\documentclass[reprint,showpacs,twocolumn,superscriptaddress,fleqn,floatfix]{revtex4-1}
\pdfoutput=1
\usepackage{graphics}
\usepackage{graphicx}
\usepackage{amsmath,amssymb,amsthm,amsfonts,epsfig}
\usepackage{bm}
\usepackage{natbib}
\usepackage[version=3]{mhchem}
\usepackage{lipsum}
\newcommand{\figref}[1]{Fig.~\ref{#1}}

\begin{document}
\title{Influence of the "second gap" on the transparency-conductivity compromise in transparent conducting oxides: an ab initio study}
\author{Viet-Anh Ha}
\affiliation{Institute of Condensed Matter and Nanoscience (IMCN), Universit\'{e} catholique de Louvain, Chemin \'{e}toiles 8, bte L7.03.01, Louvain-la-Neuve 1348, Belgium}
\author{David Waroquiers}
\affiliation{Institute of Condensed Matter and Nanoscience (IMCN), Universit\'{e} catholique de Louvain, Chemin \'{e}toiles 8, bte L7.03.01, Louvain-la-Neuve 1348, Belgium}
\author{Gian-Marco Rignanese}
\affiliation{Institute of Condensed Matter and Nanoscience (IMCN), Universit\'{e} catholique de Louvain, Chemin \'{e}toiles 8, bte L7.03.01, Louvain-la-Neuve 1348, Belgium}
\author{Geoffroy Hautier}
\email[\emph{E-mail}: ]{geoffroy.hautier@uclouvain.be}
\affiliation{Institute of Condensed Matter and Nanoscience (IMCN), Universit\'{e} catholique de Louvain, Chemin \'{e}toiles 8, bte L7.03.01, Louvain-la-Neuve 1348, Belgium}
\date{\today}
\begin{abstract}
Transparent conducting oxides (TCOs) are essential to many technologies. These materials are doped (\emph{n}- or \emph{p}-type) oxides with a large enough band gap (ideally $>$3~eV) to ensure transparency. However, the high carrier concentration present in TCOs lead additionally to the possibility for optical transitions from the occupied conduction bands to higher states for \emph{n}-type materials and from lower states to the unoccupied valence bands for \emph{p}-type TCOs. The ``second gap'' formed by these transitions might limit transparency and a large second gap has been sometimes proposed as a design criteria for high performance TCOs. Here, we study the influence of this second gap on optical absorption using \emph{ab initio} computations for several well-known \emph{n}- and \emph{p}-type TCOs. Our work demonstrates that most known \emph{n}-type TCOs do not suffer from second gap absorption in the visible even at very high carrier concentrations. On the contrary, \emph{p}-type oxides show lowering of their optical transmission for high carrier concentrations due to second gap effects. We link this dissimilarity to the different chemistries involved in \emph{n}- versus typical \emph{p}-type TCOs. Quantitatively, we show that second gap effects lead to only moderate loss of transmission (even in p-type TCOs) and suggest that a wide second gap, while beneficial, should not be considered as a needed criteria for a working TCO.
\end{abstract}
\maketitle
%

Transparent conducting materials are crucial for many technologies (e.g., displays or thin-film solar cells)\cite{H.Ohta04, K.Ellmer12}. The exceptional combination of transparency to the visible light and high conductivity can be achieved by doping large band gap oxides in order to form so-called transparent conducting oxides (TCOs). A wide variety of \emph{n}-type and \emph{p}-type doped oxides have been considered and extensively studied [e.g., \emph{n}-type: \ce{In2O3} doped with tin (ITO)\cite{I.Hamberg84, P.P.Edwards04, C.Korber10, Mryasov01, AronWalsh08} and \ce{ZnO} doped with aluminum (AZO)\cite{T.Minami85, M.A.Martinez97, H.Agura03, Y.-S.Kim06} or \emph{p}-type: \ce{SnO}\cite{Y.Ogo08, H.Yabuta10} and \ce{CuAlO2}\cite{H.Kawazoe97, J.Tate09}]. While TCOs are widespread nowadays, important efforts are still underway to discover new materials and to optimize the current ones. This is especially the case for \emph{p}-type TCOs that are lagging behind the best \emph{n}-type materials.

\begin{figure}[!ht]
\begin{center}
\includegraphics[width=0.9\linewidth]{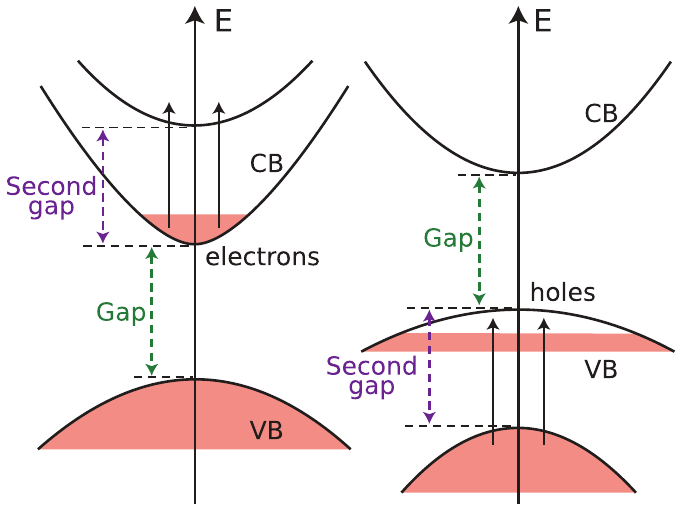}
\end{center}
\vspace{-15pt}
\caption{New transitions in highly doped TCOs. In the left panel, \emph{n}-TCOs, electronic transitions from CB to higher energy states. In the right panel, \emph{p}-TCOs, electronic transitions from lower energy states to the VB.}   
\label{secondgap}
\end{figure}

The properties of importance for a TCO are well-known (transparency, mobility, high dopability) and can be related to the fundamental electronic structure of the oxide\cite{P.P.Edwards04, G.Trimarchi11, J.B.Varley14}. This has led to an important body of computationally or chemically driven searches for good TCOs candidates\cite{GeoffroyHautier13, GeoffroyHautier14, D.O.Scanlon10, AronWalsh09, H.Kawazoe97, Y.Ogo08, H.Yabuta10, J.D.Perkins11, F.Yan15, T.F.T.Cerqueira15}. Those studies have highlighted a series of straightforward design criteria, or necessary properties for a TCO: a large band gap (ideally $>$3~eV), a low effective mass (of electrons for \emph{n}-type and holes for \emph{p}-type), and a high dopability\cite{A.Zunger03, D.O.Scanlon14, J.Robertson11}. To achieve substantial conductivity, TCOs often present carrier concentrations on the order of 10$^{21}$~cm$^{-3}$. The additional electrons or holes inserted into the conduction band(CB) (for \emph{n}-type TCOs) or the valence band(VB) (for \emph{p}-type TCOs) can lead to new optical transitions. When interacting with incident photons, the electrons in the CB of \emph{n}-type TCOs can absorb photons and undergo transitions to higher states as illustrated in the left panel of \figref{secondgap}. In a similar way, in \emph{p}-type TCOs (right panel of \figref{secondgap}), the electrons in lower states can transition up and recombine with the holes in the VB. Such second gap transitions, may affect significantly the transparency of TCOs, and therefore appear as another key criteria for the design of transparent conductors. 
%

In this paper, we investigate the influence of second gap transitions on the transparency of TCOs. The absorption coefficient is computed for various doping concentrations in various well-known \emph{n}-type (\ce{ZnO}, \ce{In2O3}, \ce{SnO2}, \ce{BaSnO3}) and \emph{p}-type TCOs (\ce{CuAlO2}, \ce{SnO}, \ce{LaCuOS}, \ce{ZnRh2O4}). We show that, at large doping concentrations, the transparency can be lowered by second gap transitions and provide a quantitative estimate of the magnitude of this effect. We also point out a fundamental dissimilarity in second gap effect in many \emph{p}-type TCOs compared to \emph{n}-types. We relate this fundamental contrast to the very different type of chemistry present in \emph{n}- versus \emph{p}-type TCOs.
%

We use the Heyd-Scuseria-Ernzherof (HSE) hybrid exchange-correlation functional as it is known to better capture the electronic structure and especially the band gap of semiconductors and insulators\cite{J.Heyd03, EdwardN.Brothers08}.
Following previous works, we use a different fraction $a$ of exact-exchange for each material in order to obtain band gaps reproducing the experimental data: $a$=25$\%$ for \ce{In2O3}\cite{AronWalsh08, AronWalsh09, PeterAgoston09}, \ce{BaSnO3}\cite{Heng-RuiLiu13}, \ce{CuAlO2}\cite{FabioTrani10, D.O.Scanlon10}, and  \ce{ZnRh2O4}\cite{D.O.Scanlon11}; $a$=32$\%$ for \ce{SnO2}\cite{J.B.Varley10} and \ce{SnO}\cite{J.B.Varley13}; and $a$=37.5$\%$ for \ce{ZnO}\cite{FumiyasuOba11}.
For \ce{LaCuOS}, we are not aware of any previous HSE studies.
So, we adopt a value of $a$=25$\%$ which leads to a good agreement with experimental data\cite{K.Ueda00} (see supplementary material). In order to study the optical properties, we calculate, at different carrier concentrations, the frequency-dependent dielectric tensor within the random-phase approximation (RPA)\cite{M.Gajdos06} by using the VASP code \cite{G.Kresse0796, G.Kresse1096} and pymatgen\cite{ShyuePingOng13} for post-treatment (more details in supplementary material).


The selected \emph{n}-type and \emph{p}-type TCOs cover a diverse range of chemistries. A series of information about these compounds (such as formula, space group, type of doping, computed and experimental direct band gap, etc.) is provided in Table~I of supplementary material.

\begin{figure*}[!ht]
\begin{center}
\includegraphics[width=0.9\linewidth]{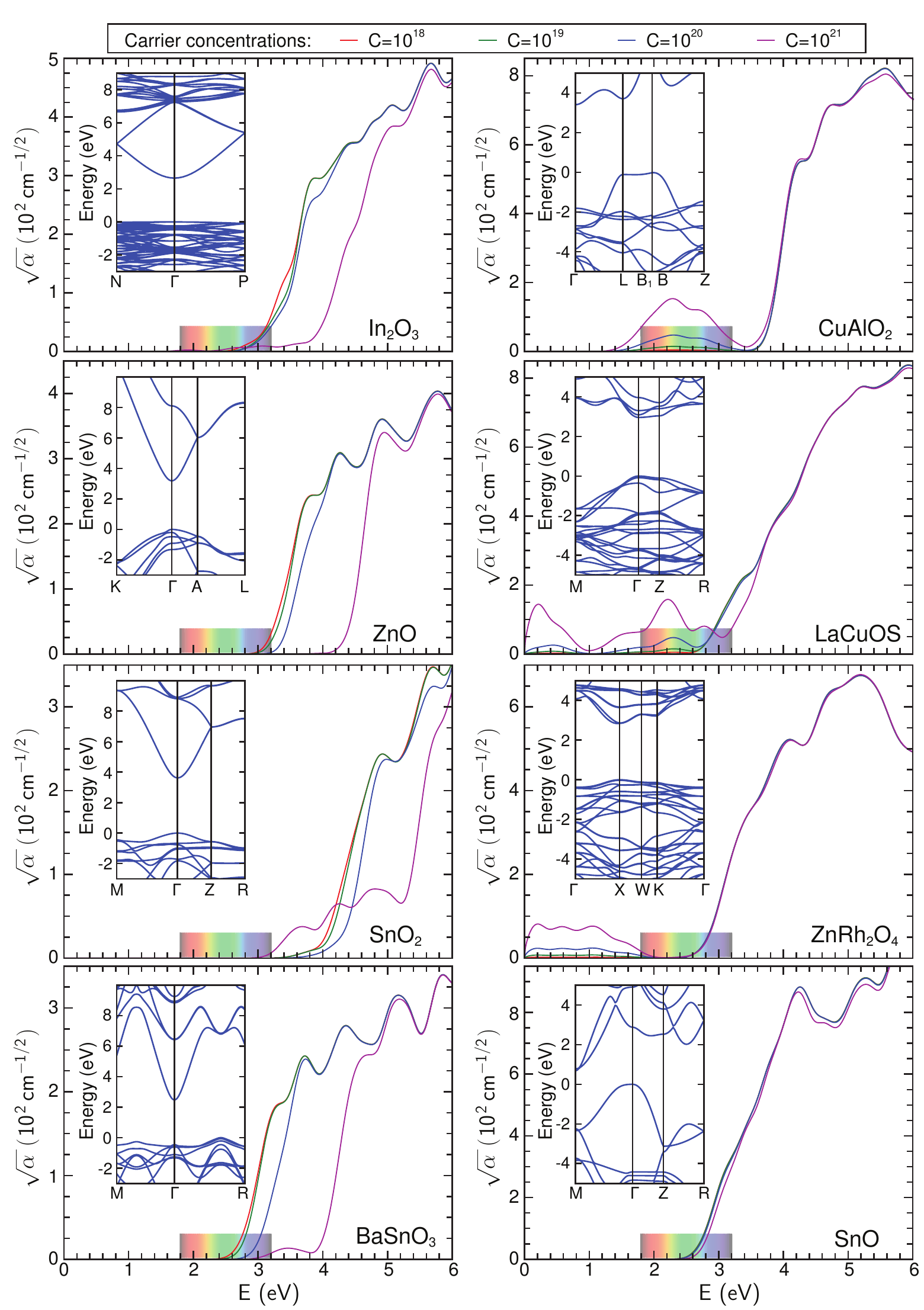}
\end{center}
\vspace{-15pt}
\caption{Square root of computed absorption coefficients of selected \emph{n}-type (left: \ce{In2O3}, \ce{ZnO}, \ce{SnO2}, and \ce{BaSnO3}) and \emph{p}-type TCOs (right: \ce{CuAlO2}, \ce{LaCuOS}, \ce{ZnRh2O4}, and \ce{SnO}) at different carrier concentrations ($C$=10$^{18}$-10$^{21}$~cm$^{-3}$). The insets are corresponding band structures. All computations are performed using the HSE hybrid functional with a fraction of exact exchange reproducing the experimental band gaps.}
\label{AbsCoePBS-np-TCO}
\end{figure*}

The computed absorption coefficients at different carrier concentrations ($C$=10$^{18}$-10$^{21}$~cm$^{-3}$) of the selected \emph{n}-type materials (\ce{In2O3}, \ce{ZnO}, \ce{SnO2}, and \ce{BaSnO3}) are reported in \figref{AbsCoePBS-np-TCO} (on the left side) together with the corresponding band structures. Interestingly, increasing the carrier concentration does not lead to any absorption in the visible range. Indeed, the second gap above the CB is quite large, typically around 4~eV, and cannot lead to contributions in the visible range. In fact, for the highest carrier concentrations ($C$$\geq$10$^{20}$~cm$^{-3}$), the optical gap for these \emph{n}-type TCOs is even widened due to the Moss-Burnstein effect\cite{MariusGrundmann06}. The right side of \figref{AbsCoePBS-np-TCO} shows the absorption for the selected \emph{p}-type materials (\ce{CuAlO2}, \ce{LaCuOS}, \ce{ZnRh2O4} and \ce{SnO}). Significant absorption appears at low energy as the hole concentration increases. The effect is particularly large for the highest hole concentrations ($C$$\geq$ 10$^{20}$~cm$^{-3}$). \ce{ZnRh2O4}, \ce{CuAlO2} and \ce{LaCuOS} all show second gap transitions at energies lower than 3~eV. Among the \emph{p}-type materials, only \ce{SnO} does not present a strong degradation of its transparency when the carrier concentration is increased. Indeed, the band structure of \ce{SnO} shows a large second gap of about 4~eV.

\begin{figure}[!ht]
\begin{center}
\includegraphics[width=0.9\linewidth]{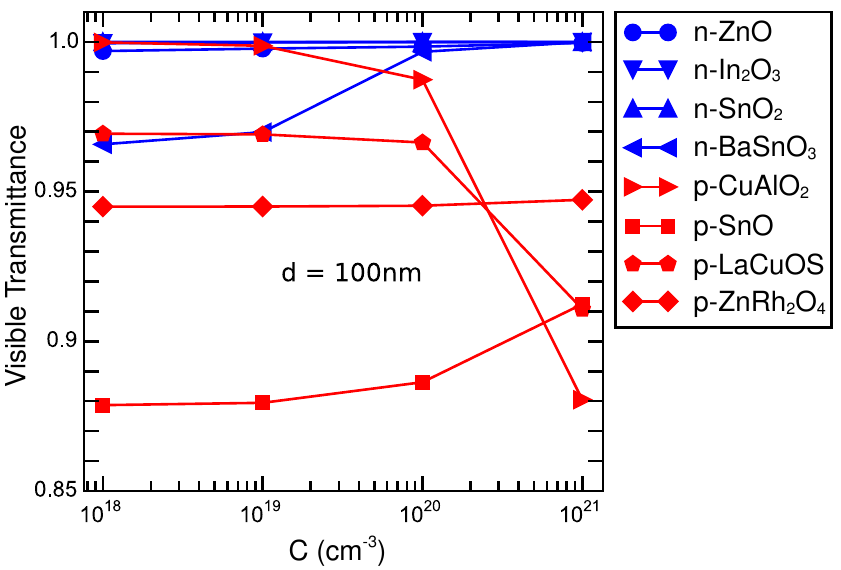}
\end{center}
\vspace{-20pt}
\caption{Visible transmittance through a 100~nm-film for the eight selected TCOs versus the carrier concentration $C$.}
\label{Transmittance}
\end{figure}

The difference between \emph{n}- and \emph{p}-type TCOs is striking. The \emph{n}-type materials tend to have large second gap transitions that do not affect transparency when doped. In contrast, most \emph{p}-type TCOs show significant absorption in the visible range due to second gap transitions with the notable exception of \ce{SnO}. A quantitative estimate of the effect of the second gap on the transparency is provided by the visible transmittance of a typical thin-film. \figref{Transmittance} plots the transmittance for a 100-nm film depending on the carrier concentration. We assume here no reflection of the incident light. The blue (red) lines correspond to \emph{n}-type (\emph{p}-type) materials. As expected from the computed absorption coefficients, \emph{n}-type TCOs do not show any degradation of their VT when the carrier concentration increases. In contrast, the transmittance of \emph{p}-type TCOs is significantly degraded. This is the case for \ce{CuAlO2} and \ce{LaCuOS}. The lower band gap of \ce{ZnRh2O4} induces a degradation in the transmission with doping but only in an energy window lower than the visible light (in the near infra-red). Finally, \ce{SnO} shows an improvement in transparency due to its large second gap combined with a Moss-Burnstein effect. We would like to stress that our analysis only takes into account direct inter-band transitions. Including also non-direct intra-band transitions~\cite{H.Peelaer12} would be significantly more computationally expensive. Moreover, we do not consider the loss of transmission due to the possible contribution of plasma reflectivity when carrier concentration is increased~\cite{K.Ellmer12, P.P.Edwards04}. Nevertheless, our study points out to lower second gap transitions as one of the reasons explaining why many \emph{p}-type TCOs show a less favorable trade-off between transparency and carrier concentration than \emph{n}-type materials do\cite{K.H.L.Zhang15}.

\begin{figure*}[!ht]
\begin{center}
\includegraphics[width=0.9\linewidth]{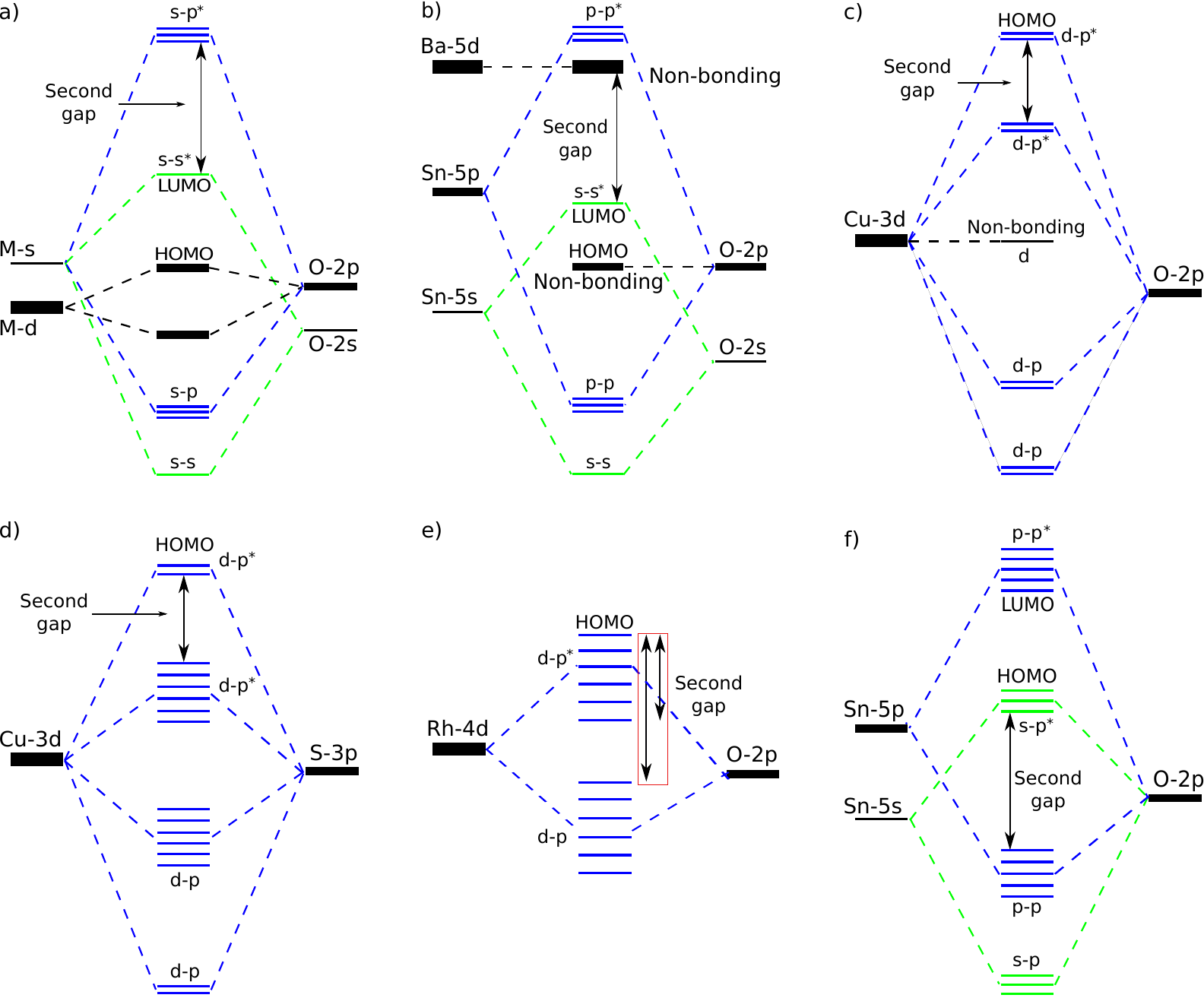}
\end{center}
\vspace{-15pt}
\caption{The chemical origin of the second gap in TCOs can be revealed by molecular orbital diagrams at particular $\boldsymbol{k}$-points of the Brillouin zone: (a) the $\Gamma$ point in \ce{In2O3}, \ce{ZnO}, and \ce{SnO2}; (b) the $\Gamma$ point in \ce{BaSnO3}; (c) the X point in \ce{CuAlO2}; (d) the $\Gamma$ point in \ce{LaCuOS}; (e) the X point in \ce{ZnRh2O4}; and (f) the $\Gamma$ point in \ce{SnO}. Bonding and anti-bonding orbitals are distinguished by marking the latter with a star. For the sake of clarity, irrelevant orbitals are not indicated.}
\label{hybridization}
\end{figure*}

The dissimilitude between \emph{n}- and \emph{p}-type TCOs can be traced back to the materials chemistry. \figref{hybridization} shows simplified molecular orbital diagrams for the studied TCOs. The corresponding projected band structures and bonding analysis data are reported in the supplementary material. In \emph{n}-type oxides (panel a), the second gap is formed by M-\emph{s}/O-\emph{s} and M-\emph{s}/O-\emph{p} anti-bonding states (with M=In, Sn or Zn) which have a large difference in energy. \ce{BaSnO3} (panel b) is slightly different with the presence of a Ba-\emph{d} non-bonding states but also provides a wide second gap.

The \emph{p}-type TCOs on the other hand (panels c to f) show more complex orbital diagrams. The materials based on transition metals (TM) have a valence band maximum (VBM) formed by TM-\emph{d}/O-\emph{p} or TM-\emph{d}/S-\emph{p} anti-bonding states (with TM=Cu or Rh). The second gap is formed between the VBM and other TM-\emph{d}/O-\emph{p}, or TM-\emph{d}/S-\emph{p} states that are very close in energy to this VBM. This leads to small second gap transitions that can be in the visible range. In contrast, \ce{SnO} presents a very different character for its VBM (Sn-\emph{s}/O-\emph{p}) and no hybridized \emph{d}-states. Hence, it shows a much larger second gap which is due to the larger hybridization of Sn-\emph{s}/O-\emph{p} states compared to \emph{d} states to O-\emph{p}. This indicates that the lower hole effective mass obtained by M-\emph{s}/O-\emph{p} hybridization is not only beneficial for the mobility of this compound but also for its second gap transitions. Our analysis points out to an intrinsic limit to transition \emph{d}-metal-based compared to \emph{s}-metal-based \emph{p}-type TCOs as they not only lead to higher effective masses but also present larger limitations in terms of second gap absorption.

We give a quantitative estimate of the effect of the second gap transitions (see \figref{Transmittance}). For a 100-nm film, the observable loss in transparency with carrier concentration reached 10$\%$ at most (20$\%$ for a 200-nm film) and for the highest carrier concentrations (10$^{20}$~cm$^{-3}$ to $10^{21}$~cm$^{-3}$). Therefore, a small second gap has a more moderate effect on transparency than the principal band gap. This comes from the smaller amount of carriers available in the small energy window corresponding to the second gap. This suggests that, while beneficial, a high second gap should not be considered a necessary condition for obtaining a working TCO.
%

In summary, we used \emph{ab initio} calculations combined with the rigid band approximation to study the effect of heavy doping on optical absorption through ``second gap'' transitions. We studied a series of common \emph{n}-type and \emph{p}-type TCOs. A very different behavior is observed between \emph{n}- and \emph{p}-type materials. The \emph{n}-type TCOs show high second gap transitions and do not lose transparency when highly doped. On the contrary, most \emph{p}-type materials have second gap transitions lower than 3~eV which lead to a degradation of their transparency when doped. We relate the asymmetry in behavior between \emph{n}- and \emph{p}-type materials to their chemistry. The lower second gap present in transition metal-based \emph{p}-type TCOs comes from the small energy differences between metal-$d/$O-$p$ hybridized states. Our results on \ce{SnO}, one of the rare non-transition metal-based \emph{p}-type TCO, indicates that using \emph{p}-type TCOs relying on the hybridization of O-\emph{p} with metal \emph{s}-states lead naturally to higher second gap transitions that are less detrimental to transparency. Our work also quantitatively estimates the effect of these second gap transitions and demonstrates that, while it is beneficial to select materials with a high second gap ($>$3~eV ideally), the magnitude of the effect should not lead to making it a strict requirement for TCO design.\\
%

V.-A.H. was funded through a grant from the FRIA. G.-M.R. is grateful to the F.R.S.-FNRS for financial support. We acknowledge access to various computational resources: the Tier-1 supercomputer of the F\'{e}d\'{e}ration Wallonie-Bruxelles funded by the Walloon Region (grant agreement N$^0$ 1117545), and all the facitilies provided by the Universit\'{e} catholique de Louvain (CISM/UCL) and by the Consortium des \'{E}quipements de Calcul Intensif en F\'{e}d\'{e}ration Wallonie Bruxelles (C\'{E}CI). The authors thank Professor Chris Van de Walle for helpful discussions.

%
%
\bibliographystyle{apsrev4-1}
\bibliography{biblio}
\end{document}


\title{Supplementary Material for\\Influence of the "second gap" on the transparency-conductivity compromise in transparent conducting oxides: an \emph{ab initio} study}
\author{Viet-Anh Ha}
\affiliation{Institute of Condensed Matter and Nanoscience (IMCN), Universit\'{e} catholique de Louvain, Chemin \'{e}toiles 8, bte L7.03.01, Louvain-la-Neuve 1348, Belgium}
\author{David Waroquiers}
\affiliation{Institute of Condensed Matter and Nanoscience (IMCN), Universit\'{e} catholique de Louvain, Chemin \'{e}toiles 8, bte L7.03.01, Louvain-la-Neuve 1348, Belgium}
\author{Gian-Marco Rignanese}
\affiliation{Institute of Condensed Matter and Nanoscience (IMCN), Universit\'{e} catholique de Louvain, Chemin \'{e}toiles 8, bte L7.03.01, Louvain-la-Neuve 1348, Belgium}
\author{Geoffroy Hautier}
\email[\emph{E-mail}: ]{geoffroy.hautier@uclouvain.be}
\affiliation{Institute of Condensed Matter and Nanoscience (IMCN), Universit\'{e} catholique de Louvain, Chemin \'{e}toiles 8, bte L7.03.01, Louvain-la-Neuve 1348, Belgium}
\date{\today}
\maketitle
\section{COMPUTATIONAL DETAILS}
We use the Heyd-Scuseria-Ernzheroff (HSE) hybrid exchange-correlation functional \cite{J.Heyd03, EdwardN.Brothers08}. Following previous works, we use a different fraction $a$ of exact-exchange for each material in order to obtain band gaps reproducing the experimental data: $a$=25$\%$ for \ce{In2O3}\cite{AronWalsh08, AronWalsh09, PeterAgoston09}, \ce{BaSnO3}\cite{Heng-RuiLiu13}, \ce{CuAlO2}\cite{FabioTrani10, D.O.Scanlon10}, and  \ce{ZnRh2O4}\cite{D.O.Scanlon11}; $a$=32$\%$ for \ce{SnO2}\cite{J.B.Varley10} and \ce{SnO}\cite{J.B.Varley13}; and $a$=37.5$\%$ for \ce{ZnO}\cite{FumiyasuOba11}. For \ce{LaCuOS}, we are not aware of any previous HSE studies. So, we adopt a value of $a$=25$\%$ which leads to a good agreement with experimental data\cite{K.Ueda00} (more details are given in the \tabref{table.I}). All the calculations are performed using the VASP code\cite{G.Kresse0796, G.Kresse1096}. The densest $\boldsymbol{k}$-point mesh used in calculations of optical absorption is up to $17 \times 17 \times 17$. Then, in order to study the optical properties, we calculate the frequency-dependent dielectric tensor within the random-phase approximation (RPA) at different carrier concentrations\cite{M.Gajdos06}. The formulas and relevant quantities are developed hereafter.

The imaginary part of the dielectric tensor is given by:
\begin{align}
\varepsilon_{\alpha \beta}^{(2)}(\omega) = & \frac{4\pi^2e^2}{\Omega} \lim_{q \rightarrow 0} \frac{1}{q^2} \sum_{n, n', \boldsymbol{k}} 2w_{\boldsymbol{k}} \delta(\epsilon_{n\boldsymbol{k}}-\epsilon_{n'\boldsymbol{k}}-\omega)[f(\epsilon_{n'\boldsymbol{k}})-f(\epsilon_{n\boldsymbol{k}})]\langle u_{n\boldsymbol{k}+\boldsymbol{q}_{\alpha}} | u_{n'\boldsymbol{k}} \rangle \langle u_{n\boldsymbol{k}+\boldsymbol{q}_{\beta}} | u_{n'\boldsymbol{k}} \rangle^{\ast},
\label{imaginaryDT} 
\end{align}
where $\alpha$ and $\beta$ refer to the three space directions, $\omega$ is the photon frequency, $e$ is the elementary charge, $\Omega$ is the volume of the cell, $q$ is the incident photon wave vector, $w_{\boldsymbol{k}}$ are the weights of the different $\boldsymbol{k}$-points in the irreducible Brillouin zone, $\epsilon_{n\boldsymbol{k}}$ is the energy of $n^{th}$ eigenenergy at $\boldsymbol{k}$-point, $|u_{n\boldsymbol{k}} \rangle$ are the Bloch wave functions, and
\begin{equation}
f(\epsilon) = \left(1 + e^{(\epsilon - E_f)/k_BT}\right)^{-1}
\label{FermiDis}
\end{equation}
is the Fermi-Dirac distribution at the temperature $T$. Here, $E_f$ is the Fermi level, and $k_B$ is the Boltzmann constant.
The real part of the dielectric tensor can be found through the Kramers-Kronig relation:
\begin{equation}
\varepsilon_{\alpha \beta}^{(1)}(\omega) = 1 + \frac{2}{\pi} \mathcal{P}\int_0^{\infty} \frac{\varepsilon_{\alpha \beta}^{(2)}(\omega') \omega'}{\omega'^2 - \omega^2}d\omega',
\label{realDT}
\end{equation}
where $\mathcal{P}$ denotes the Cauchy principal value\cite{Vladimirov71}.

We introduce the effect of doping by varying the Fermi level so as to reach a certain carrier concentration using the following formulas:
\begin{align}
& n = \int_{CBM}^{\infty}f(E)g(E)dE, \vspace{2mm}\\
& p = \int_{-\infty}^{VBM}[1-f(E)]g(E)dE,
\label{concentration}
\end{align}
where $E$ is the energy, $f(E)$ is the Fermi-Dirac distribution as defined in Eq.~\eqref{FermiDis}, and $g(E)$ is the density of state (DOS).
We adopt the rigid band approximation which assumes that the doping does not modify the band structure and only plays a role in the Fermi level. The matrix elements:
\begin{equation}
\lim_{q \rightarrow 0} \frac{1}{q^2} \langle u_{n\boldsymbol{k}+\boldsymbol{e_{\alpha}}q} | u_{n'\boldsymbol{k}} \rangle \langle u_{n\boldsymbol{k}+\boldsymbol{e_{\beta}}q} | u_{n'\boldsymbol{k}} \rangle^{\ast}
\end{equation}
are determined from the DFT calculations. The dielectric tensor is computed for various Fermi levels at 300K using Eqs. \eqref{imaginaryDT} and \eqref{realDT}. Our implementation relies on computations of the matrix elements within an ab initio code (here, VASP) and post-processing including the carrier concentration dependence using the python-based pymatgen code\cite{ShyuePingOng13}.

The optical absorption coefficient is derived from the dielectric tensor as follows\cite{Claudia06}: 
\begin{equation}
\alpha(\omega) = \frac{2\omega k(\omega)}{c}
\label{absorptionCoeff}
\end{equation}
where $c$ is the speed of light and $k(\omega)$ is the extinction coefficient defined by:
\begin{equation}
k(\omega) = \sqrt{\frac{\sqrt{\varepsilon_1^2(\omega) + \varepsilon_2^2(\omega)}-\varepsilon_1(\omega)}{2}}.
\end{equation}
In this formula, $\varepsilon_1(\omega)$ and $\varepsilon_2(\omega)$ are determined as the average of the three diagonal components of the dielectric tensors $\varepsilon_{\alpha \beta}^{(1)}(\omega)$ and $\varepsilon_{\alpha \beta}^{(2)}(\omega)$, respectively.
We should note that computations have been performed recently on various perovskites to also obtain the dependence of carrier concentrations dependence of optical absorption but taking into account the specific dopant specie in a virtual crystal approximation framework.\cite{YuweiLi15, KhuongP.Ong15}. Our approach does not model the type of dopant inserted and only assumes a rigid band approach and doping setting a given Fermi level (and carrier concentration). Please note that our approach does not take into account indirect transitions (involving phonons) as well as the increase in reflectivity due to plasmons and present in highly doped semiconductors.
\section{MATERIALS DETAILS}
We have selected a series of well-known representative \emph{n}-type and \emph{p}-type TCOs covering a diverse range of chemistries. We have considered \ce{In2O3}, \ce{ZnO}, \ce{SnO2} and \ce{BaSnO3} as \emph{n}-type and \ce{CuAlO2}, \ce{LaCuOS}, \ce{ZnRh2O4} and \ce{SnO} as \emph{p}-type TCOs. Table \tabref{table.I} provides their chemical formula, their space group, their identification number (MP-id) in the Materials Project\cite{MatPro} database in which all the details about their structures can be found, the type of doping considered, the fraction of exact exchange $a$ used in the HSE calculations, their computed direct band gap $E^\textrm{HSE}_{dg}$, and as well as their reported experimental direct band gap $E^\textrm{exp}_{dg}$. The chosen values for the parameter $a$ guarantee a fair consistence between $E^\textrm{HSE}_{dg}$ and $E^\textrm{exp}_{dg}$.
\begin{table*}[!ht]
\caption{Formula, space group, Materials Project\cite{MatPro} identification number (MP-id), type of doping, fraction of exact exchange ($a$), computed direct band gap ($E^\textrm{HSE}_{dg}$), and experimental direct band gap ($E^\textrm{exp}_{dg}$) for all TCOs studied in this work.}
\begin{center}
\renewcommand{\arraystretch}{1.3}
\begin{tabular*} {0.7\textwidth} {@{\extracolsep{\fill}} l l r c l c c}
\hline
Formula      & Space group  & MP-id  & Doping & $a$(\%) & $E^\textrm{HSE}_{dg}$(eV) & $E^\textrm{exp}_{dg}$(eV) \\
\hline
\ce{In2O3}   & $Ia3$        & 22598  & \emph{n}     & 25      & 2.65           & 2.9\cite{AronWalsh08} \\
\ce{ZnO}     & $P6_3mc$     & 2133   & \emph{n}     & 37.5    & 3.19           & 3.3\cite{U.Ozgur05}  \\
\ce{SnO2}    & $P4_2/mnm$   & 856    & \emph{n}     & 32      & 3.62           & 3.5\cite{M.Nagasawa66} \\
\ce{BaSnO3}  & $Pm\bar{3}m$ & 3163   & \emph{n}     & 25      & 2.91           & 3.1\cite{HyungJKim12} \\
\ce{CuAlO2}  & $R\bar{3}m$  & 3748   & \emph{p}     & 25      & 3.42           & 3.47\cite{J.Tate09} \\
\ce{SnO}     & $P4/nmm$     & 2097   & \emph{p}     & 32      & 2.89           & 2.7\cite{Y.Ogo08} \\
\ce{LaCuOS}  & $P4/nmm$     & 6088   & \emph{p}     & 25      & 2.95           & 3.1\cite{K.Ueda00} \\
\ce{ZnRh2O4} & $Fd\bar{3}m$ & 5146   & \emph{p}     & 25      & 2.84           & 2.74\cite{M.Dekkers07} \\
\hline   
\end{tabular*}
\label{table.I}
\end{center}
\end{table*}
\section{Chemical rules for the second gap}
Our analysis of the orbital hybridization and band structure character for the TCO candidates is based on projected DFT band structures computed with VASP~\cite{G.Kresse0796, G.Kresse1096} within the PAW method. We use both standard atomic projected band structures and Crystal Orbital Hamilton Populations (COHP) analysis~\cite{R.Dronskowski93}. The COHP are computed using the Lobster package~\cite{V.L.Deringer11, S.Maintz13}. The set of basis functions is chosen to guarantee a small charge spilling (ideally $<$1\%).

\figref{COHP-PB-nTCO} presents the atomically projected band structure and the COHP of the studied \emph{n}-TCOs. Three compounds \ce{In2O3}, \ce{ZnO}, \ce{SnO2} contain metals with an (n-1)$d^{10}$n$s^{2}$ configuration (\ce{Zn}: n=3 and \ce{In}, \ce{Sn}: n=4). Their projected band structure and COHP show that, at the $\Gamma$ point, the orbital M-$s$ of metals and O-2$s$ form the CBM through an anti-bonding interaction. The higher energy band state above the CBM defining the second gap comes from anti-bonding M-$s/$O-2$p$ states. Our observations are consistent with previous works~\cite{J.L.G.Fierro06, A.Facchetti10}. A similar trend is observed at the $\Gamma$ point of \ce{BaSnO3} but with some differences. A Ba-5$d$ non-bonding is present between the CBM and the Sn-5$p/$O-2$p$ anti-bonding states. Note that this differs from a report of Mizoguchi~\emph{et al.}\cite{Mizoguchi04}. We have constructed Figure 4 in the main text based on this information.
\begin{figure*}[!ht]
\begin{center}
\includegraphics[width=0.9\linewidth]{Figures/fig01.pdf}
\end{center}
\vspace{-15pt}
\caption{Projected DFT band structure [left] and Crystal Orbital Hamilton Populations (COHP) [right] of selected \emph{n}-TCOs.}
\label{COHP-PB-nTCO}
\end{figure*}

\figref{COHP-PB-pTCO} shows the atomically projected band structure and COHP analysis for our selected p-type TCOs. 
\begin{figure*}[!ht]
\begin{center}
\includegraphics[width=0.9\linewidth]{Figures/fig04.pdf}
\end{center}
\vspace{-15pt}
\caption{Projected DFT band structure [left] and Crystal Orbital Hamilton Populations (COHP) [right] of selected \emph{p}-TCOs.}
\label{COHP-PB-pTCO}
\end{figure*}
\section{Visible transmittance}
In order to provide a quantitative estimate of the effect of the second gap transitions, we compute the visible transmittance, which is defined as the fraction of energy in the visible range which passes through a thin film of given thickness. It is obtained by the following formula:
\begin{equation}
T_{V} = \frac{\int_{1.8}^{3.2} e^{-\alpha d} I(\omega) d\omega}{\int_{1.8}^{3.2} I(\omega) d\omega},
\end{equation}
where $\alpha$ and $d$ are the absorption coefficient and the thickness of the thin film, respectively; and, $I(\omega)$ is the distribution of the radiation spectrum with $\omega$ is photon energy.
We assume that the incident light is perpendicular to the surface of thin film and that there is no reflection. We consider a uniform radiation distribution within the visible range (1.8-3.2~eV).

\begin{figure*}[!ht]
\begin{center}
\includegraphics[width=0.9\linewidth]{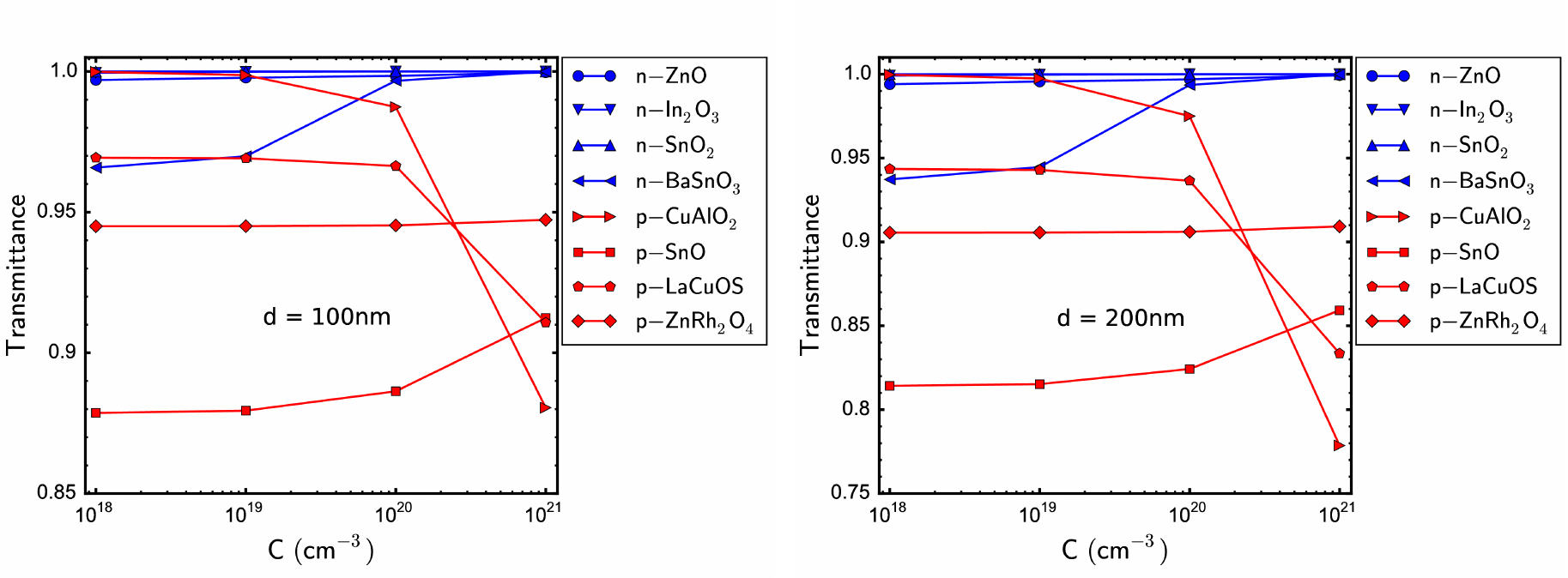}
\end{center}
\vspace{-15pt}
\caption{Visible transmittance through 100-nm (left) and 200-nm (right) films for the eight selected TCOs as a function of the carrier concentration $C$.}
\label{Transmittance}
\end{figure*}

\figref{Transmittance} shows $T_{V}$ for two thin films of 100~nm and 200~nm as the function of doping concentration. For the 100-nm (resp. 200-nm) film, the observable loss in transparency with carrier concentration reaches 10$\%$ (resp. 20$\%$) at most. 
\bibliographystyle{apsrev4-1}
\bibliography{supbiblio}